\documentclass[lettersize, journal]{IEEEtran}
\usepackage{amsmath, amssymb, amsfonts, amsbsy, amsthm, mathtools}
\usepackage{mathrsfs, dsfont, bbm}
\usepackage{nicefrac}
\usepackage{microtype}
\usepackage{soul}
\usepackage{afterpage}
\usepackage{balance}

\usepackage{graphicx}
\usepackage{wrapfig}
\usepackage{booktabs}
\usepackage{adjustbox} 
\usepackage{multirow}
\usepackage{makecell}
\usepackage{xcolor}
\usepackage{nicematrix}
\usepackage{tikz}
\usepackage{caption}    
\usepackage{subfig}

\usepackage{cite}
\def\BibTeX{{\rm B\kern-.05em{\sc i\kern-.025em b}\kern-.08em
    T\kern-.1667em\lower.7ex\hbox{E}\kern-.125emX}}

\usepackage{url}

\theoremstyle{plain}

\usepackage{balance}

\def\gap{0.75ex}
\abovedisplayskip\gap
\belowdisplayskip\gap
\abovedisplayshortskip\gap
\belowdisplayshortskip\gap

\begin{document}

\title{Few-Shot Beam Selection for Embodied AI via In-Context Learning and Conformal Prediction}

\author{Weicao Deng, Ruihan Li, Feihang Xu, Min Li
\thanks{
The work was supported in part by a Project Supported by Scientific Research Fund of Zhejiang University (No. XY2025015).

The authors are with the College of Information Science and Electronic Engineering, the Zhejiang Provincial Key Laboratory of Multi-Modal Communication Networks and Intelligent Information, and the National Key Laboratory of Millimeter-Wave and Terahertz Remote Sensing, Zhejiang University, Hangzhou 310027, China (e-mail: \{caowd, ruihan.li, feihang.xu, min.li\}@zju.edu.cn).}}

\maketitle
\thispagestyle{empty}
\begin{abstract}
Millimeter-wave (mmWave) communications are essential for supporting massive multimodal sensory data exchange in Embodied Artificial Intelligence (EAI) systems. However, fast and reliable beam selection for mmWave communication remains a critical bottleneck due to the prohibitive overhead associated with conventional beam training.
Although location-aided beam prediction offers a promising alternative, its performance is often limited by localization errors and the need for extensive training data.
To address these challenges, we propose a novel framework termed In-Context Learning driven Cross-Validation Conformal Prediction (ICL+CVCP) for beam prediction. 
The proposed framework utilizes the training-free adaptation capability of ICL, enabling effective operation with only a small number of demonstration samples and eliminating the need for large-scale data collection and model fine-tuning.
In addition, the CVCP module constructs compact, distribution-free candidate beam sets that provably contain the optimal beam with a user-defined probability. 
Experimental results confirm that ICL+CVCP consistently satisfies the desired empirical coverage while significantly reducing pilot overhead and computational latency, thereby meeting the stringent reliability and efficiency requirements of EAI applications.
\end{abstract}


\section{Introduction}
\label{sec:introduction}
The recent paradigm shift toward Embodied Artificial Intelligence (EAI) has accelerated the development of autonomous systems, including intelligent robotics and uncrewed aerial vehicles, which actively interact with complex physical environments \cite{cangelosi2015embodied}. 
To support the offloading of massive multimodal sensory data (e.g., high-resolution video and LiDAR point clouds) and enable real-time control, EAI devices require ultra-reliable and high-capacity wireless communication \cite{li2025large}. 
Millimeter-wave (mmWave) systems have emerged as key enabler for meeting these stringent data rate requirements. 
However, due to severe path loss at high frequencies, mmWave communication relies heavily on directional beamforming \cite{8751142,8454665,11168837}. 
Consequently, fast and reliable beam selection becomes critical for ensuring robust connectivity in EAI applications.

Despite its importance, efficient beam selection remains challenging. 
Traditional beam selection, such as exhaustive sweeping and hierarchical searching, incur substantial latency, which limits the real-time responsiveness of EAI systems \cite{11071925,11226846}. 
Alternatively, location-aided methods utilize position information (e.g., GPS or SLAM) to reduce the search space. 
However, the deployment of these methods in EAI scenarios face several challenges. 
First, strict privacy constraints often prevent EAI devices from sharing their precise location information with the access point (AP)\cite{zhu2024romans}, while the AP may not disclose its deployment details due to commercial confidentiality and the prevention of targeted physical-layer attacks\cite{jian2007protecting}. 
Second, EAI devices typically operate in dynamic environments where collecting large-scale training datasets for model training is costly or infeasible.
These factors lead to significant information asymmetry and data scarcity, placing the system in a \textit{few-shot} regime where only limited measurements are available for beam selection.

Under such conditions, localization errors \cite{torrieri2007statistical} and incomplete information sharing significantly degrade the reliability of beam selection.
Although existing heuristic schemes \cite{alrabeiah2020deep,ma2022deep} can generate candidate beam sets, they lack formal, distribution-free reliability guarantees, leaving EAI systems vulnerable to link failures.
To address these limitations, conformal prediction (CP) and conformal risk control \cite{kuchibhotla2020exchangeability} provide a principled framework for uncertainty quantification with distribution-free guarantees, and have demonstrated effectiveness across various domains \cite{11077387,angelopoulos2023conformal,10584135,zhang2026prediction}. 
Nevertheless, their application to wireless physical layer designs remains largely underexplored \cite{10262367,deng2025scan}.
In particular, the application of CP to few-shot beam selection under scarce and environment-specific data remains a significant open challenge.

To bridge this gap, In-Context Learning (ICL) offers a promising paradigm for rapid task adaptation \cite{chen2023learning}. 
By leveraging the pattern-matching capabilities of foundation models, ICL enables accurate prediction from few examples without explicit parameter updates. 
Inspired by this, recent works \cite{huangpredictions,11267082} have explored applying ICL to full conformal prediction (FCP).
However, these approaches incur prohibitive computational complexity in beam selection over high-dimensional label spaces, as they require exhaustive evaluation of all test sample-candidate label pairs.

In this context, we propose a novel, location-aided beam selection framework termed \textit{In-Context Learning driven Cross-Validation Conformal Prediction (ICL+CVCP)}. 
Specifically, in the proposed framework, ICL enables efficient adaptation to new environments using minimal data, eliminating the need for costly model fine-tuning.
Building upon this, CVCP, more efficient in high-dimensional label spaces, converts point predictions into compact candidate beam sets with rigorous, distribution-free reliability guarantees, ensuring that the optimal beam is included with a user-defined probability.
Extensive experimental results demonstrate that ICL+CVCP consistently satisfies target coverage requirements and achieves high optimal beam hit rates. 
Moreover, by producing significantly tighter candidate sets, the proposed approach substantially reduces pilot probing overhead. Combined with its low computational complexity, ICL+CVCP provides a robust and low-latency solution for beam selection in dynamic EAI systems.

\section{System Model and Problem Definition}
\subsection{Location-Aided Beam Selection for EAI Devices}
\begin{figure}
    \centering
    \includegraphics[width=0.95\linewidth]{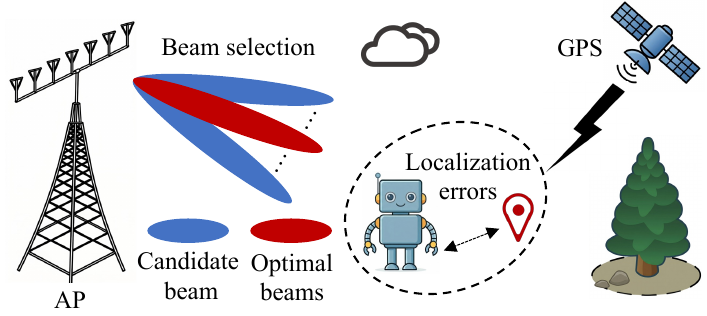}
    \caption{System model of location-aided beam selection for EAI devices.}
    \label{sys_model}
\end{figure}
As illustrated in Fig. \ref{sys_model}, we consider a narrowband mmWave communication system operating at a carrier frequency $f_{\text{c}}$ with bandwidth $B$. The access point (AP) is equipped with an $N_{\text{t}}$-antenna uniform linear array (ULA) with half-wavelength spacing. Let $\mathbf{x}_0=(a_0,b_0,c_0)$ and $\theta$ denote the unknown 3D coordinates and the array orientation of the AP, respectively. 

To avoid the prohibitive pilot overhead of exhaustive beam sweeping, the EAI device (equipped with a single antenna) leverages its own 3D GPS coordinates $\mathbf{x}=(a,b,c)$ as prior spatial information, thereby explicitly exploiting the geometry-dependent nature of mmWave channels \cite{sun2017millimeter}.
However, practical GPS measurements are subject to inherent localization errors. Furthermore, due to strict privacy regulations, neither the AP nor the EAI device shares their absolute coordinate information or hardware specifications with the other \cite{zhu2024romans}.

Let $\mathbf{h}^T \in \mathbb{C}^{1 \times N_{\text{t}}}$ denote the channel vector from the AP to the EAI device. The AP employs a pure analog beamforming architecture, where the beamforming vector $\mathbf{f} \in \mathbb{C}^{N_{\text{t}} \times 1}$ is selected from a predefined Discrete Fourier Transform (DFT) codebook $\mathcal{W} \triangleq \{\mathbf{w}_1, \dots, \mathbf{w}_{N_\text{c}}\}$. Without loss of generality, we assume $N_\text{c} = N_{\text{t}}$. For a given channel $\mathbf{h}^T$, the optimal beam index $y \in \{1, \dots, N_\text{c}\}$, is chosen to maximize the received signal gain, i.e., 
\begin{equation}
    y = \mathop{\arg\max}_{1 \leqslant j \leqslant N_{\text{c}}} \left\| \mathbf{h}^T \mathbf{w}_j \right\|_2^2.
\end{equation}
Accordingly, the optimal beam is $\mathbf{f}^\star = \mathbf{w}_y$. 
Given a noisy position observation $\mathbf{x} \in \mathbb{R}^3$, the posterior probability distribution over the optimal beam index is denoted as $p(\cdot|\mathbf{x})$.

\subsubsection*{Few-Shot Setting}
In unseen environments, the EAI device encounters inevitable domain shifts due to unknown AP configurations and varying GPS noise statistics. Under such conditions, collecting a large-scale labeled dataset is practically prohibitive. Consequently, the system is constrained to a \textit{few-shot} regime, where only a limited support set of $n$ labeled location-beam samples, denoted as $\mathcal{D} = \{(\mathbf{x}_i, y_i)\}_{i=1}^{n}$, is available to estimate the posterior distribution $\hat{p}(\cdot|\mathbf{x})$.

\subsection{Uncertainty Quantification and Problem Formulation}

In the few-shot regime, the estimated posterior distribution $\hat{p}(y|\mathbf{x})$ is subject to significant uncertainty, arising from both noisy GPS observations and the extreme scarcity of environmental samples. 
Consequently, point prediction strategies, such as greedily selecting the beam with the maximum predictive probability, are insufficient to ensure reliable connectivity for EAI devices. To address this limiation, instead of outputting a single beam, we aim to construct a candidate beam set $\Gamma(\mathbf{x}_{\text{te}})$ for a given test location $\mathbf{x}_{\text{te}}$, followed by further limited beam training to identify the final beam $\mathbf{f}$.

The efficacy of the constructed candidate set $\Gamma(\mathbf{x}_{\text{te}})$ hinges on an inherent trade-off between two critical metrics: \textit{coverage} and \textit{inefficiency}. \emph{Coverage} is defined as the marginal probability that the optimal beam is contained within the candidate set, i.e., $\mathbb{P}(y_{\text{te}} \in \Gamma(\mathbf{x}_{\text{te}}))$, which essentially determines the ultimate reliability of the established mmWave link.
In particular, when beam training is restricted to this set, the coverage closely approximates the probability that the final selected beam $\mathbf{f}$ is optimal.
Conversely, \emph{inefficiency} is quantified by the expected size of the candidate set, $\mathbb{E}[|\Gamma(\mathbf{x}_{\text{te}})|]$, which reflects the pilot overhead required for beam training.

To balance this trade-off, our objective is to design an efficient location-aided beam selection framework that constructs a compact candidate set $\Gamma_{\alpha}(\mathbf{x}_{\text{te}})$ while providing formal reliability guarantees. Specifically, for a user-defined miscoverage level $\alpha \in (0,1)$ (e.g., $\alpha=0.1$ for a $90\%$ reliability target), the generated set must strictly satisfy the following marginal coverage condition:
\begin{equation}
\mathbb{P}\big(y_{\text{te}} \in \Gamma_{\alpha}(\mathbf{x}_{\text{te}})\big) \geqslant 1 - \alpha. \label{eq:objective}
\end{equation}

\section{ICL+CVCP: In-Context Learning-Driven Cross-Validation Conformal Prediction}
In this section, we first introduce the classical CVCP for formal coverage, and subsequently propose the training-free ICL+CVCP scheme for efficient few-shot beam coverage guarantees.

\subsection{Classical Cross-Validation Conformal Prediction}
\label{subsec:cvcp}
As a powerful tool for uncertainty quantification, CVCP improves data utilization by partitioning the dataset $\mathcal{D}$ into $K$ subsets. 
While conventional $K$-fold approximations ($K \ll n$) are computationally efficient, they compromise sample exchangeability and thus lack strict theoretical guarantees for the marginal coverage in \eqref{eq:objective} \cite{barber2021predictive}.
In the extreme data scarcity regime of few-shot scenarios, the leave-one-out (LOO) setting ($K=n$) is therefore preferred, as it maximizes data utilization while preserving exchangeability and formal coverage guarantees \cite{barber2021predictive}.

Specifically, each sample $(\mathbf{x}_i,y_i)$ is sequentially treated as a calibration sample, while the remaining $n-1$ samples (i.e., $\mathcal{D} \setminus \{(\mathbf{x}_i,y_i)\}$) form the corresponding training set to learn the $i$-th prediction model $\hat{p}^{(-i)}(\cdot|\mathbf{x})$. Based on this predictor, the non-conformity score for the calibration sample $(\mathbf{x}_i,y_i)$ is calculated as
\begin{equation}
    s_i = -\log\left(\hat{p}^{(-i)}(y_{i}|\mathbf{x}_{i})\right). \label{eq:calibration_sample_NCS}
\end{equation}
After iterating through all $n$ samples in $\mathcal{D}$, the calibration score set $\{s_i\}_{i=1}^{n}$ is obtained.

Next, for a hypothesized sample $(\mathbf{x}_{n+1},y)$ formed by a test sample $\mathbf{x}_{n+1}$ and any candidate label $y \in \mathcal{Y}$, all $n$ predictors $\{\hat{p}^{(-i)}(\cdot|\mathbf{x})\}_{i=1}^n$ are utilized to calculate their respective non-conformity scores
\begin{equation}
    s_{i,n+1}^{y} = -\log\left(\hat{p}^{(-i)}(y|\mathbf{x}_{n+1})\right), \label{eq:test_sample_NCS_y}
\end{equation}
and the minimum non-conformity score for $(\mathbf{x}_{n+1},y)$ is computed as
\begin{equation}
    s_{\text{min},n+1}^y = \min_{i=1,\dots,n} s_{i,n+1}^y. \label{eq:eq:test_sample_NCS_y_min}
\end{equation}
By comparing this score with the calibration score set $\{s_i\}_{i=1}^n$, the $p$-value $\pi(y)$ for the candidate label $y$ is calculated as
\begin{equation}
    \pi(y) = \frac{1}{n+1}\sum_{i=1}^n \mathds{1}(s_{\text{min},n+1}^y \leqslant s_i).
    \label{eq:cvcp_p_value}
\end{equation}
Accordingly, the prediction set for the test input $\mathbf{x}_{n+1}$ is defined as
\begin{equation}
    \Gamma_{\alpha}(\mathbf{x}_{n+1}) = \{\,y \in \mathcal{Y}: \pi(y) \geqslant \alpha \,\}.
    \label{eq:cvcp_final_prediction_set}
\end{equation}

Despite its strong theoretical guarantees, classical CVCP faces significant practical challenges in few-shot settings. First, training on extremely small LOO subsets increases the risk of severe overfitting. Second, the requirement of retraining n separate models incurs substantial computational overhead, limiting its applicability in resource-constrained EAI systems.
\subsection{In-Context Learning}
\label{subsubsec:icl}
Unlike conventional meta-learning methods that rely on iterative parameter updates, 
ICL requires no explicit gradient fine-tuning of the model; it can rapidly adapt to new tasks simply by implicitly extracting patterns from context samples during forward inference \cite{chen2023learning}. Specifically, given a context set $\mathcal{D}_{\text{con}} = \{(\mathbf{x}_i, y_i)\}_{i=1}^{n}$ and a query input $\mathbf{x}_{n+1}$, the ICL model $G_{\boldsymbol{\theta}}(\cdot,\cdot)$ parameterized by $\boldsymbol{\theta}$ directly outputs the predictive distribution corresponding to $\mathbf{x}_{n+1}$:
\begin{equation}
    \hat{p}(\cdot|\mathbf{x}_{n+1}) = G_{\boldsymbol{\theta}}(\mathcal{D}_{\text{con}}, \mathbf{x}_{n+1}).
    \label{eq:conventional_ICL_model}
\end{equation}
From a probabilistic perspective, this output distribution $\hat{p}(\cdot|\mathbf{x}_{n+1})$ serves as an effective approximation to the posterior distribution for $y_{n+1}$ given $\mathcal{D}_{\text{con}}$ and $\mathbf{x}_{n+1}$, leveraging the model's learned prior knowledge \cite{xie2021explanation}.

To train $G_{\boldsymbol{\theta}}(\cdot,\cdot)$, data are first collected from $T$ auxiliary tasks \cite{wies2023learnability}. For each task $t=1,\dots,T$, a context sample set $\mathcal{D}_{\text{con}}^{t}=\{(\mathbf{x}_i^t, y_i^t)\}_{i=1}^n$ and a query sample pair $(\mathbf{x}_{n+1}^t, y_{n+1}^t)$ are drawn independently and identically distributed (i.i.d.) from a task-specific distribution.The model parameters $\boldsymbol{\theta}$ are then optimized by minimizing the negative log-likelihood loss across all tasks:
\begin{equation}
    L(\boldsymbol{\theta}) = -\frac{1}{T}\sum_{t=1}^{T} \log\left(G_{\boldsymbol{\theta}}(\mathcal{D}_{\text{con}}^{t},\mathbf{x}_{n+1}^{t})\right).
    \label{eq:classical_pretrained_loss}
\end{equation}
Minimizing $L(\boldsymbol{\theta})$ encourages the model to assign high probability to the true query label $y_{n+1}^t$ across all auxiliary tasks, thereby enabling it to infer task-specific mappings directly from the provided context $\mathcal{D}_{\text{con}}^{t}$.

\subsection{ICL+CVCP}
\label{subsec:icl_cvcp}
\begin{figure}[t]
    \centering
    \includegraphics[width=1.0\linewidth]{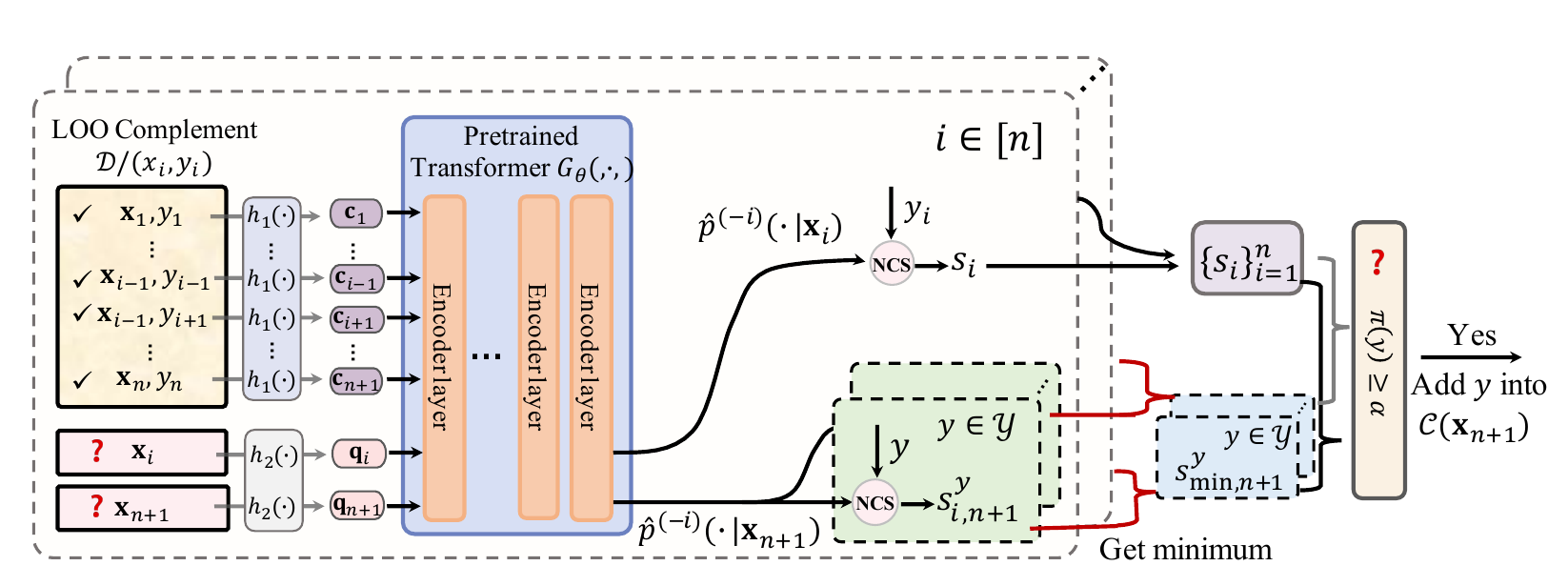} 
    \caption{Schematic diagram of the proposed ICL+CVCP. By paralleling Transformer~$G_{\boldsymbol{\theta}}(\cdot,\cdot)$ to emulate the predictors~$\{\hat{p}^{(-i)}(\cdot|\mathbf{x})\}_{i=1}^n$, the framework eliminates the repeated retraining operations required in conventional CVCP.}
    \label{fig:ICL+CVCP}
\end{figure}

As illustrated in Fig.~\ref{fig:ICL+CVCP}, for each sample $(\mathbf{x}_i, y_i)$ in the dataset $\mathcal{D}=\{(\mathbf{x}_i,y_i)\}_{i=1}^n$, a LOO complement set $\mathcal{D} \setminus \{(\mathbf{x}_i,y_i)\}$ is constructed to serve as the context for prediction. 
This complement set is mapped by an embedding network $h_1(\cdot)$ into a sequence of context features $\{\mathbf{c}_1,\dots,\mathbf{c}_{i-1},\mathbf{c}_{i+1},\dots,\mathbf{c}_n\}$. Simultaneously, the current calibration input $\mathbf{x}_i$ and the test input $\mathbf{x}_{n+1}$ are encoded by another embedding network $h_2(\cdot)$ into query features $\mathbf{q}_i$ and $\mathbf{q}_{n+1}$, respectively. 
Subsequently, these features are concatenated into a unified input sequence $[\mathbf{c}_1,\dots,\mathbf{c}_{i-1},\mathbf{c}_{i+1},\dots,\mathbf{c}_n,\mathbf{q}_i,\mathbf{q}_{n+1}]$ , which is then fed into a Transformer model with cascaded encoder layers.

To ensure the permutation invariance required by CVCP, the attention mask matrix $\mathbf{M}_{\text{cv}}$ is designed as \cite{fang2025rethinking}
\begin{equation}
\mathbf{M}_{\text{cv}} =
{\scriptsize
\begin{bNiceMatrix}[first-row,first-col]
    & \text{Context Features} & \text{Query Features} \\
\text{Context Features} 
    & \Block{1-1}{\mathbf{0}_{(n-1)\times(n-1)}} 
    & \Block{1-1}{-\infty \cdot \mathbf{1}_{(n-1)\times 2}} \\
\text{Query Features}   
    & \Block{1-1}{\mathbf{0}_{2\times(n-1)}} 
    & \Block{1-1}{-\infty(\mathbf{1}_{2\times 2}-\mathbf{I}_{2})}
\end{bNiceMatrix},}\label{eq:M_cvcp}
\end{equation}
where we adopt the convention $\infty \cdot 0 = 0$. 
This mask enforces that context features interact globally, context features attend globally to each other, while query features attend only to the context and themselves. 
Thus, the final two encoder outputs correspond to the predictive distributions $\hat{p}^{(-i)}(\cdot|\mathbf{x}_{i})$ and $\hat{p}^{(-i)}(\cdot|\mathbf{x}_{n+1})$, respectively.

By performing $n$ such forward inferences in parallel, the prediction results $\{\hat{p}^{(-i)}(\cdot|\mathbf{x}_{i})\}_{i=1}^n$ and $\{\hat{p}^{(-i)}(\cdot|\mathbf{x}_{n+1})\}_{i=1}^n$ can be efficiently obtained for all LOO complement sets. This design eliminates the need for repeated model retraining, thereby avoiding the substantial computational overhead of classical CVCP.

Using the resultant predictive distributions $\{\hat{p}^{(-i)}(\cdot|\mathbf{x}_{i})\}_{i=1}^n$, a global calibration score set $\{s_i\}_{i=1}^n$ is constructed by calculating the non-conformity score (NCS) for each sample in $\mathcal{D}$ via \eqref{eq:calibration_sample_NCS}. 
Subsequently, for a test input $\mathbf{x}_{n+1}$ and a hypothesized label $y \in \mathcal{Y}$, the corresponding test NCS and its minimum are evaluated according to \eqref{eq:test_sample_NCS_y} and \eqref{eq:eq:test_sample_NCS_y_min}, respectively. 
The $p$-value for label $y$ is then obtained via \eqref{eq:cvcp_p_value}, yielding the final prediction set $\Gamma_{\alpha}(\mathbf{x}_{n+1})$ as defined in \eqref{eq:cvcp_final_prediction_set}.

\paragraph*{Built-in KV Cache Acceleration}
ICL+CVCP naturally leverages the Key-Value (KV) Cache mechanism of large language models. Since the LOO context features $[\mathbf{c}_1,\dots,\mathbf{c}_{i-1},\mathbf{c}_{i+1},\dots,\mathbf{c}_n]$ and query $\mathbf{q}_i$ remain fixed across different new test samples $\mathbf{x}_{n+1}$, their corresponding KV pairs are computed only once. Reusing these cached representations eliminates redundant forward passes, drastically reducing computational overhead without compromising accuracy.

\subsection{Limited Beam Training}
\label{subsec:limited_beam_training}
Based on the candidate set $\Gamma_{\alpha}(\mathbf{x})$ generated by ICL+CVCP, the system adopts an adaptive, limited beam training and selection strategy:
\begin{itemize}
    \item \textbf{Limited pilot measurement:} If $\Gamma_{\alpha}(\mathbf{x})$ contains multiple codeword indices ($|\Gamma_{\alpha}(\mathbf{x})| > 1$), an uplink beam training is performed only over the candidate beams in $\Gamma_{\alpha}(\mathbf{x})$ to determine the final downlink beam $\mathbf{f}$:
    \begin{equation}
        \mathbf{f}=\mathbf{w}_{j^*}, \quad j^*=\mathop{\arg\max}_{j\in \Gamma_{\alpha}(\mathbf{x})} |\mathbf{w}_j^T\mathbf{h}s_{\text{p}}+\mathbf{w}_j^T\mathbf{n}|^2,
    \end{equation}
    where $\mathbf{w}_j$ denotes the $j$-th codeword, $\mathbf{h}$ is the uplink channel vector, $s_{\text{p}}=\sqrt{P_{\text{u}}}$ is the user pilot signal with transmit power $P_{\text{u}}$, and $\mathbf{n}\sim \mathcal{CN}(\mathbf{0}, \sigma_{\text{n}}^2\mathbf{I}_{N_{\text{t}}})$ is the Gaussian noise vector with power $\sigma_{\text{n}}^2$.
    
    \item \textbf{Training free selection:} If $\Gamma_{\alpha}(\mathbf{x})$ contains exactly one index ($|\Gamma_{\alpha}(\mathbf{x})| = 1$), the codeword $\mathbf{w}_j$ is directly selected as $\mathbf{f}$, thereby eliminating the need for beam training.
    
    \item \textbf{Fallback mechanism:} If $\Gamma_{\alpha}(\mathbf{x})$ is empty indicating that no reliable candidate satisfies the prescribed confidence level, a fallback strategy is triggered. In this case, the codeword associated with the label having the maximum $p$-value as $\mathbf{f}$.
\end{itemize}

\section{Results and Discussion}
\begin{figure*}[t]
\centering 
\subfloat[$1-\alpha=0.7$]{\includegraphics[width=0.33\textwidth]{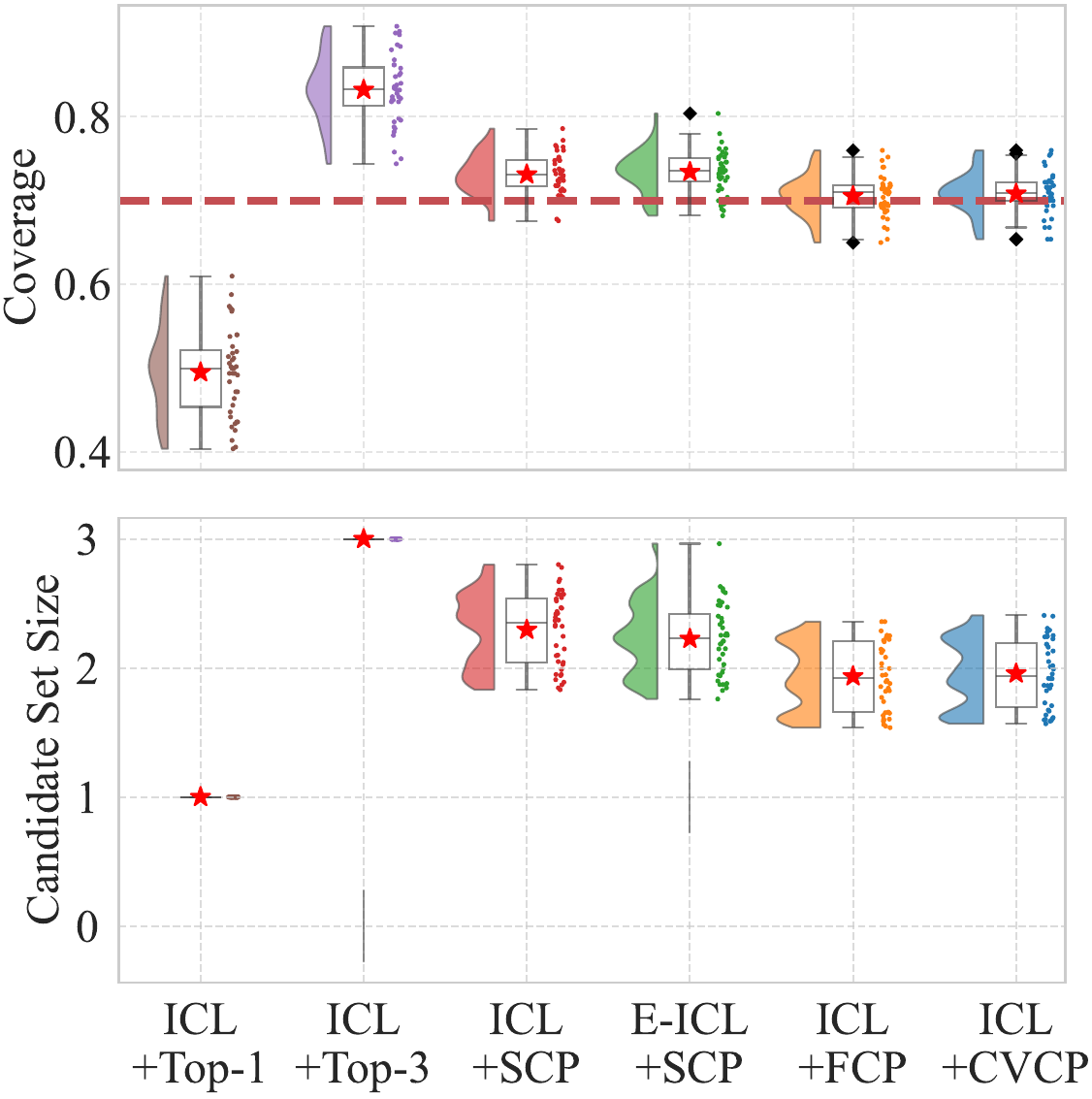}}\label{fig:coverage_and_size:0.7}
\subfloat[$1-\alpha=0.8$]{\includegraphics[width=0.33\textwidth]{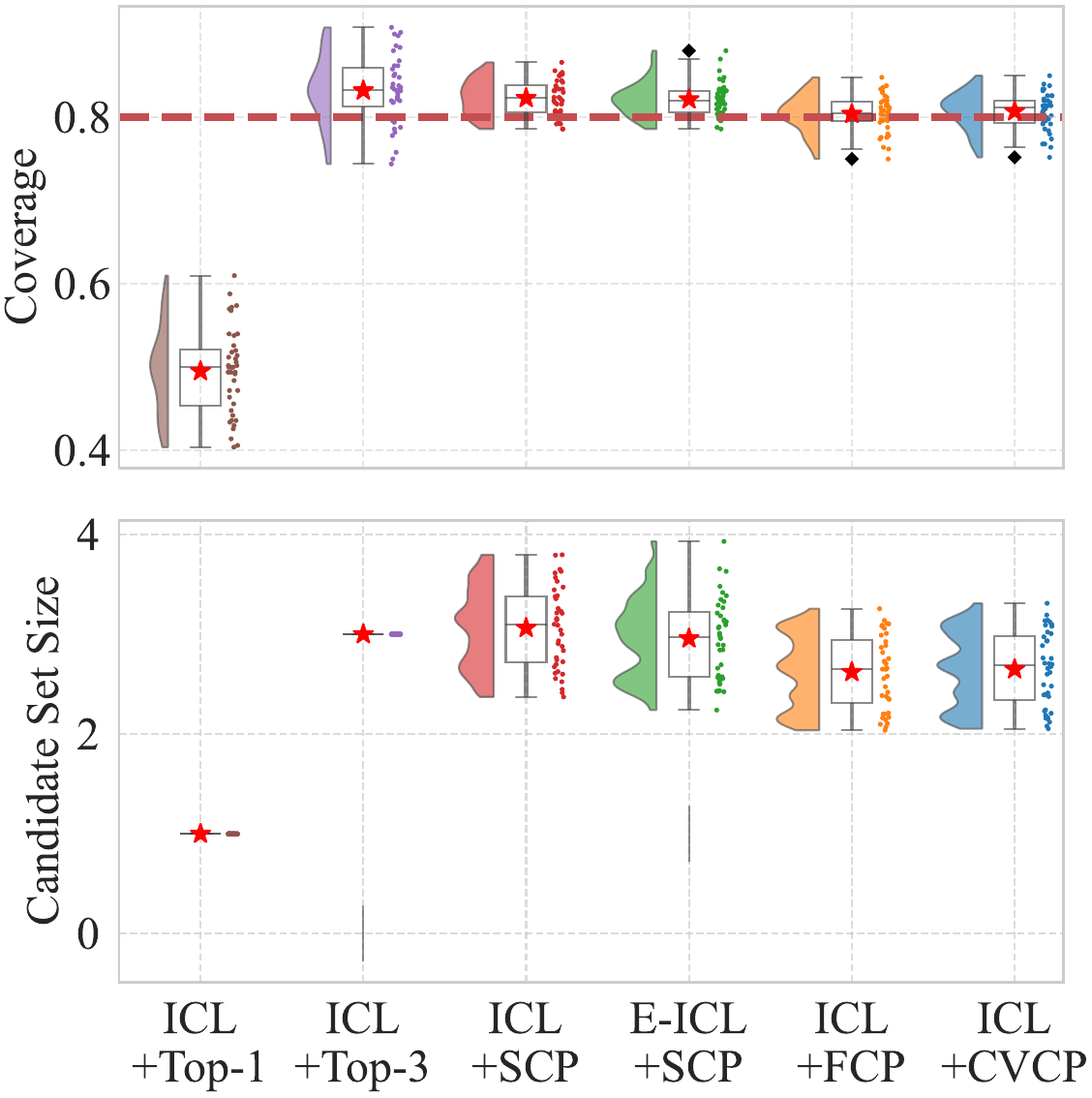}\label{fig:coverage_and_size:0.8}}
\subfloat[$1-\alpha=0.9$]{\includegraphics[width=0.33\textwidth]{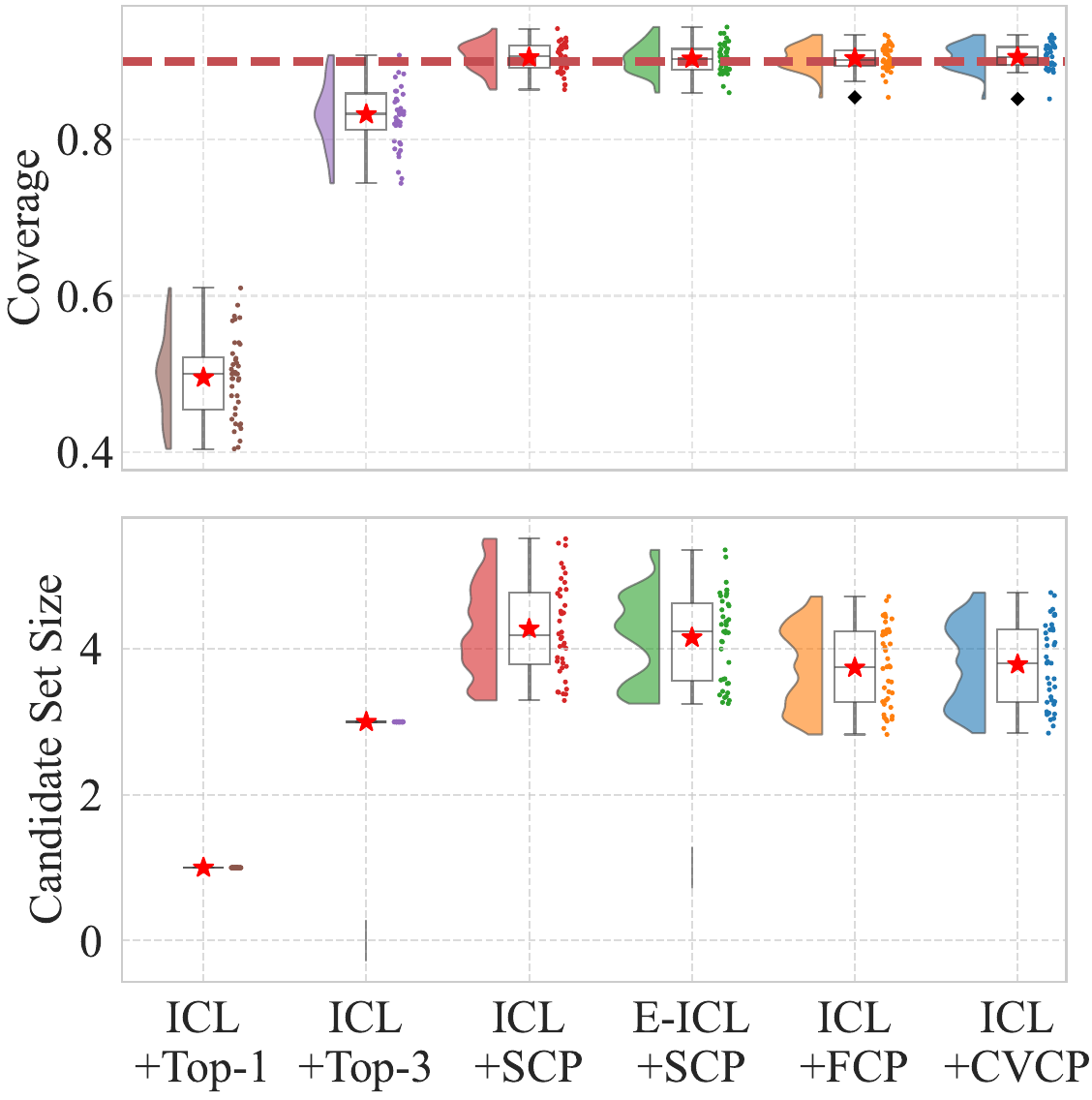}}\label{fig:coverage_and_size:0.9}
\caption{Box plots for the coverage and candidate set size under different target coverage rates $1-\alpha$ of (a) 0.70, (b) 0.80, and (c) 0.90. The red lines denote the target coverage rates $1-\alpha$, while $\textcolor{red}{\bigstar}$ marks the mean.}
\label{fig:coverage_and_size}
\end{figure*}
In this section, we evaluate the performance of ICL+CVCP in terms of coverage and inefficiency, and compare it with existing benchmark methods.

\subsection{Simulation Settings and Training Parameters}

\paragraph*{Simulation Parameters}
The channel data is generated within a $100\text{ m} \times 100\text{ m}$ area using the QuaDRiGa simulator \cite{jaeckel2014quadriga}. 
The system operates at a carrier frequency of $f_{\text{c}} = 28$ GHz with a bandwidth of $B = 20\,\text{MHz}$. 
The AP is deployed at a height of 10 m and equipped with $N_{\text{t}} = 128$ antennas, corresponding to a codebook size of $N_{\text{c}} = 128$. 
Its horizontal locations are uniformly distributed within a radius of 20m around the center of area, and its array orientation follows from a uniform distribution, $\theta \sim \mathcal{U}(0, 2\pi)$.
The EAI device is positioned at a height of 1.5 m, with an uplink pilot power of $P_{\text{u}} = 23$ dBm, unless otherwise specified.
The horizontal GPS error of EAI device is modeled as a zero-mean Gaussian random variable with standard deviation $\sigma \sim \mathcal{U}(0.5, 3.0)$ m.
The noise power during uplink beam training is given by $\sigma_{\text{n}}^2\,(\text{dBm}) = -173.8 + 10\log_{10}(B)$.
Under these settings, each task is characterized by a specific parameter tuple $(a_0,b_0,\theta,\sigma)$.

\paragraph*{Network Architecture and Training Settings}
The ICL model is implemented as a 6-layer Transformer encoder with an input dimension of 256, 4 attention heads, and a FFN dimension of 1024. The embedding functions $h_1(\cdot)$ and $h_2(\cdot)$ are realized using linear layers. 
To train and evaluate the model, we generate 120 tasks for meta-training, 40 tasks for validation, and 40 tasks for testing.
For each task, the available dataset $\mathcal{D}$ contains $n=19$ samples. We sample 50 independent and identically distributed (i.i.d.) realizations of $\mathcal{D}$ per task, evaluating 10 i.i.d. test queries $\mathbf{x}_{n+1}$ per realization. All models are trained for 200 epochs with an initial learning rate of $2\times 10^{-4}$, utilizing the CosineAnnealingLR scheduler \cite{pytorch_cosineannealinglr_docs} configured with a maximum of 50 iterations and a minimum learning rate of $2\times 10^{-5}$.

\subsection{Baseline Methods}
Conventional joint learning approaches consistently exhibit inferior performance, while model-agnostic meta-learning suffers from suboptimal accuracy and substantial fine-tuning overhead in few-shot scenarios \cite{brown2020language}. Therefore, we adopt ICL as the common predictive backbone for all learning-based baselines.
To comprehensively validate the proposed ICL+CVCP framework, we focus on two key aspects: candidate set efficiency and pilot overhead reduction. 

First, to demonstrate its effectiveness in generating compact and reliable candidate sets under few-shot conditions, we compare it against the following four learning-based baselines:
\begin{itemize}
    \item \textbf{ICL+SCP:} Based on ICL model described in \ref{subsubsec:icl}, this method randomly partitions the dataset into separate context and calibration subsets. NCSs are computed on the calibration set to determine a rejection threshold for constructing the candidate set.
    \item \textbf{E-ICL+SCP:} ICL model trained with CP-aware loss \cite{stutz2021learning}, while retaining the same inference procedure as ICL+SCP.
    \item \textbf{ICL+FCP \cite{huangpredictions}:} Built on ICL model in Sec. \ref{subsubsec:icl}, this method iteratively augments the context with each test input-label hypothesis and reference samples. It computes NCSs via repeated inference to determine candidate inclusion. Although FCP typically yields the tightest prediction sets, it incurs substantial computational overhead.
    \item \textbf{ICL+Top-$L$:} This baseline constructs the candidate set by selecting the top-$L$ beams with the highest predicted probabilities from the ICL model as described in Sec. \ref{subsubsec:icl}.
\end{itemize}

In addition, to evaluate potential reduction in mmWave pilot overhead, we also include two conventional beam sweeping schemes:
\begin{itemize}
    \item \textbf{Exhaustive search (ES):} This scheme sequentially probes all $N_c$ beams in the codebook. While it provides an upper bound on beam selection performance, it incurs prohibitive pilot overhead.
    
    \item \textbf{Hierarchical search (HS):} This scheme employs a multi-resolution, tree-structured codebook to progressively refine beam alignment from coarse to fine levels. However, it is prone to error propagation, particularly when early-stage decisions are inaccurate.
\end{itemize}

\subsection{Coverage Guarantees and Inefficiency of ICL+CVCP}
Fig.~\ref{fig:coverage_and_size} presents the coverage and inefficiency performance of ICL+CVCP under different target reliability levels.
The results show that the CP-based methods satisfy the prescribed coverage targets by adaptively adjusting the size of the candidate sets, whereas ICL+Top-$L$ provides only fixed coverage.
Notably, the proposed ICL+CVCP constructs significantly more compact candidate sets than E-ICL+SCP and ICL+SCP, owing to its more efficient utilization of limited data.
For example, at $1-\alpha=0.90$, ICL+CVCP reduces the average candidate set size by 11.42\% and 8.84\% compared to ICL+SCP and E-ICL+SCP, respectively. 
Importantly, its performance closely approaches that of the computationally intensive ICL+FCP baseline, incurring only a marginal 1.20\% increase in set size while avoiding the associated exhaustive computation.

\subsection{Optimal Beam Hit Rate and Pilot Overhead Analysis}
\begin{figure}
    \centering
    \includegraphics[width=1.0\linewidth]{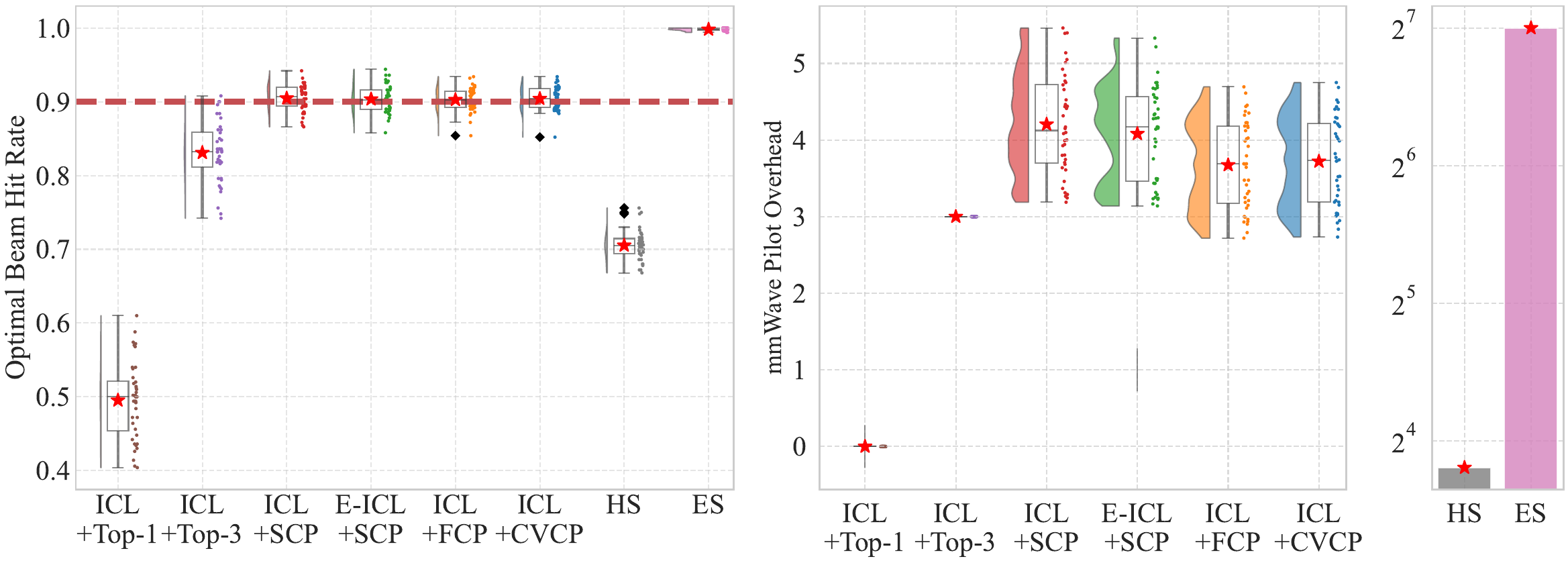}
    \caption{Optimal beam hit rate and pilot overhead of ICL+CVCP and baselines, when the coverage target is $1-\alpha=0.90$.}
    \label{fig:true optimal rate and pilot}
\end{figure}
Fig.~\ref{fig:true optimal rate and pilot} compares the optimal beam hit rate and the corresponding pilot overhead for ICL-CVCP and baseline methods at a target coverage of $1-\alpha=0.9$. 
The results indicate that localization-aided methods consistently achieve a superior trade-off between beam alignment accuracy and pilot overhead, significantly outperforming conventional ES and HS schemes. 
Furthermore, all CP-based methods strictly satisfy the predefined hit rate target, demonstrating that the theoretical, distribution-free guarantees of conformal prediction translate into reliable performance in practical wireless environments. 
Among these methods, ICL+CVCP achieves pilot overhead comparable to that of ICL+FCP, while reducing overhead by 11.50\% and 8.84\% relative to the ICL+SCP and E-ICL+SCP, respectively.

\subsection{Computational Complexity Analysis}
Table~\ref{tab:mmWave_beam_selection_complexity} summarizes the theoretical number of forward evaluations and the actual inference runtime ($r=10$) during the calibration phase, measured on an NVIDIA RTX 4090 GPU and an Intel Xeon Gold 6240 CPU.
The proposed ICL+CVCP demonstrates a clear efficiency advantage over ICL+FCP.
CVCP avoids the exhaustive enumeration of all candidate labels required by FCP, reducing the number of evaluations from $|\mathcal{Y}|r(2n+2)$ to $K(n+r)$. 
Furthermore, at the inference level, CVCP maintains a fixed reference context when processing a given input, enabling effective reuse of KV cache in Transformer models. 
In contrast, FCP repeatedly reconstructs the context, preventing such reuse.
As a result, ICL+CVCP achieves a substantial reduction in runtime compared to ICL+FCP (5.840 ms vs. 69.650 ms), while maintaining performance close to the lightweight ICL+Top-$L$ baseline. 
Overall, ICL+CVCP trades a negligible loss in candidate set compactness for a significant gain in computational efficiency, effectively overcoming the scalability limitations of conformal prediction in few-shot EAI beam alignment.
\begin{table}[t]
	\centering
	\caption{Computational Complexity of Different Methods ($r$ denotes the number of test inputs). Runtime is measured at $r=10$.}
	\label{tab:mmWave_beam_selection_complexity}
	\linespread{1.5}\selectfont
	\setlength{\tabcolsep}{2pt}
	\scriptsize
	\begin{tabular}{c|c c c|c|c} 
		\toprule
		&\makecell{ICL+Top-$L$} & \makecell{ICL+SCP} & \makecell{E-ICL+SCP} 
		& \makecell{ICL+FCP} & \makecell{ICL+CVCP} \\
		\midrule
		\# Evaluations     
		& \multicolumn{3}{c|}{$n+r$} 
		& $|\mathcal{Y}|r(2n+2)$& $K(n+r)$\\
		\makecell{Runtime (ms)}
		&5.124
		&5.115
		&5.236
		&69.650
		&5.840\\
		\bottomrule
	\end{tabular}
\end{table}

\section{Conclusions}
In this paper, we have presented ICL+CVCP, an efficient location-aided beam selection framework for EAI systems in mmWave networks. 
To address data scarcity and localization errors, the proposed framework leverages ICL for rapid, training-free adaptation from few-shot demonstrations for beam prediction. By integrating CVCP, the framework transforms unreliable point predictions into compact candidate sets with rigorous, distribution-free reliability guarantees.
Comprehensive evaluations confirm that ICL+CVCP consistently satisfies user-defined coverage requirement and achieve high optimal beam hit rates, while producing significantly tighter candidate sets than existing baselines.
Moreover, the proposed framework substantially reduces both pilot probing overhead and computational complexity. Overall, ICL+CVCP offers a robust beam selection solution, making it well-suited for dynamic and resource-constrained EAI applications.

\balance
\bibliographystyle{IEEEtran}
\bibliography{IEEEabrv,ref}
\end{document}